\documentclass[10pt]{article}
\usepackage[OE]{express}
\newcommand{\p}{\partial}

\begin{document}

\title{Spectral wings of the fiber supercontinuum  and the dark-bright soliton interaction}

\author{C. Mili\'an,\authormark{1,2} T. Marest,\authormark{3} A. Kudlinski,\authormark{3} and D.V.~ Skryabin\authormark{4,5,*}}

\address{\authormark{1}Centre de Physique Th\'{e}orique, CNRS, \'{E}cole Polytechnique, F-91128 Palaiseau, France\\
\authormark{2}ICFO--Institut de Ciencies Fotoniques, The Barcelona Institute of Science and Technology, 08860 Castelldefels (Barcelona), Spain\\
\authormark{3}{Universit\'e Lille, CNRS, UMR 8523--PhLAM--Physique des Lasers Atomes et Mol\'ecules, F-59000 Lille, France}\\
\authormark{4}Department of Physics, University of Bath, Bath BA2 7AY, UK\\
\authormark{5}Department of Nanophotonics and Metamaterials, ITMO University, St Petersburg, Russia\\
\authormark{*}d.v.skryabin@bath.ac.uk}

\begin{abstract} We present experimental and numerical data on the supercontinuum generation in an optical fiber pumped in the normal dispersion range where the seeded dark and the spontaneously generated bright solitons contribute to the spectral broadening. We report  the dispersive radiation arising from the interaction of the bright and dark solitons. This radiation consists of the two weak dispersing pulses that continuously shift their frequencies and  shape the short and long wavelength wings of the supercontinuum spectrum. 
\end{abstract}


\section*{Introduction}
A multitude of nonlinear processes  occurring in the supercontinuum generation in optical fibers is depended on the pump conditions and fiber properties, but practically in all cases, it  is strongly associated with the optical solitons spontaneously generated from the either pulsed or cw inputs \cite{DudleyRMP,SkryabinRMP}. Typically, an intense pump is launched in a spectral range with the relatively small and anomalous group velocity dispersion (GVD), so that it breaks up into  multiple bright solitons. The subsequent soliton dynamics is  strongly influenced by the Raman effect and emission of dispersive (Cherenkov) radiation  \cite{DudleyRMP,SkryabinRMP}. The research into the role the dark solitons can play in nonlinear wave mixing and supercontinuum generation has remained for a long confined to several theoretical studies \cite{Karp,Afan,MilianOL,OreshnikovOL}. However, ourselves and our collaborators have recently put this work onto the experimental footing by demonstrating supercontinuum generation by a train of dark solitons propagating in the normal GVD range and emitting an intense Cherenkov continuum into the anomalous GVD range  \cite{MarestOL}. The related research into optical shock waves and their role in supercontinuum generation has been also  pursued  recently, see, e.g., \cite{ConfortiOL}, as a part of a broader effort into understanding of the supercontinuum generation scenarios in fibers pumped in the normal GVD range, see, e.g., \cite{normal,normal2,normal3}. 

In this work, we present a supercontinuum generation scheme in optical fibers involving interacting  bright and dark solitons. We demonstrate that this interaction results in a new type of  radiation that appears in the normal and anomalous GVD ranges and shapes the long and short wavelength edges of the continuum that are both continuously shifting with propagation and are noticeably detuned from the central and most intense part of the continuum spectrum.
\section*{Results}
We proceed by briefly describing the equations used  to numerically model and interpret experimental measurements of the supercontinuum generation. An evolution of the complex envelope $A$ of the electric field inside a fiber is described with the help of the  well established generalized nonlinear Schr$\ddot{\mathrm{o}}$dinger  equation \cite{DudleyRMP,SkryabinRMP}
\begin{eqnarray}
&& i\partial_z A+\hat D(\p_t)A+\gamma|A|^2A+QA=0 \label{eq:fiber}, \
\end{eqnarray}
where $\hat D(\p_t)$ is the  operator reproducing the dispersion profile of the fiber over the entire span of the generated spectra \cite{DudleyRMP,SkryabinRMP}. $z$ and $t$  are the  coordinate along the fiber length and the time in a reference frame moving with the pump pulse group velocity. $\gamma$ is the nonlinear fiber parameter \cite{AgrawalBOOK}. $Q$ accounts for the Raman effect standard for the silica fibers  \cite{DudleyRMP,SkryabinRMP}.
In our experiments we excite the fiber using two input pulses that are identical, but delayed one with respect to the other, so that the initial envelope $A$ takes the form
\begin{equation}
A(t,z=0)=\sqrt{P_0}\left[\exp\left\{-\left(\frac{t+t_{del}}{\sqrt{2}T_0}\right)^2\right\}+\exp\left\{-\left(\frac{t}{\sqrt{2}T_0}\right)^2\right\}\right],\label{eq:Ain}
\end{equation}
where $t_{del}$ is the delay, $T_0=\tau_0/1.665$, and $\tau_0$ is the full width at half maximum (FWHM) of the transform limited pulses. Two pulse initial conditions and the input frequency in the range of the normal dispersion of the fiber were used to initiate the interference pattern between the  two dispersing pulses and subsequent generation of a train of dark solitons according to the well known method \cite{RothenbergOC,MilianOL,MarestOL}. 
\begin{figure}
\centering
\includegraphics[width=\textwidth]{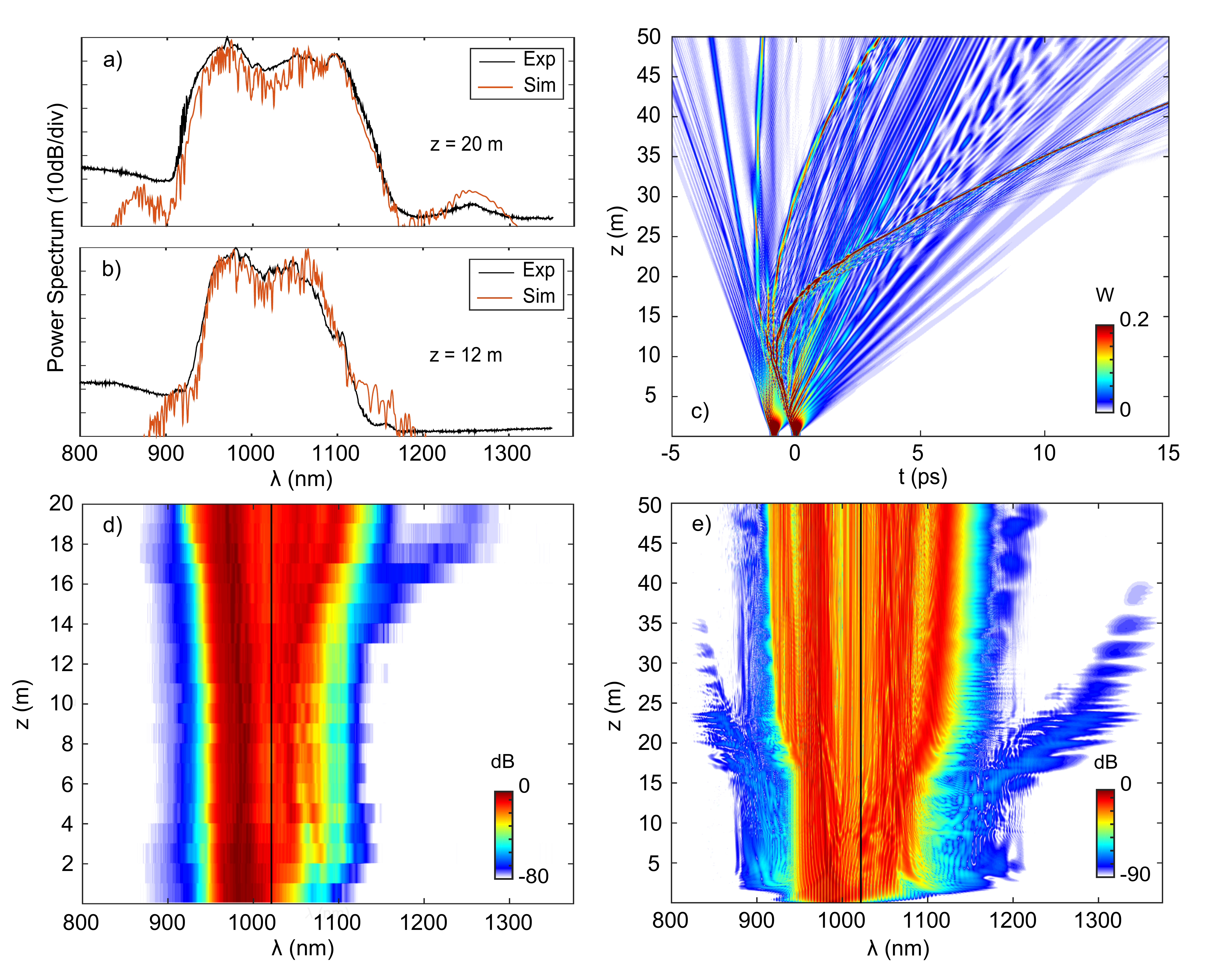}
\caption{Experimental measurements and numerical modeling of the spontaneous formation of bright solitons on top of the train of dark solitons. (a,b) Comparison of the generated spectra and of the corresponding modeling results for two fiber lengths $z=12$ and $20$m. (c) shows the simulated time domain dynamics.
(d) is the measured spectral evolution long the fiber by the cut-back technique. (e) is the numerical modeling corresponding to (d), but for longer distances. The peaks around $870$ and $1250$ nm in (b,d,e) correspond to the continuum wings associated with the interaction between the bright and dark solitons. Parameters of the input pulses and of the fibers are described in the text. \label{f2}}
\end{figure}

Figure \ref{f2} shows the experimental  \ref{f2}(a), \ref{f2}(b), \ref{f2}(d) and numerical \ref{f2}(a)--\ref{f2}(c), \ref{f2}(e) data describing  propagation of the two sufficiently intense pump pulses in a photonic crystal fiber.  The input parameters are  $P_0=685$ W, $\tau_0 =137$ fs, $t_{del} =0.88$ ps, as per Eq. (\ref{eq:Ain}), and the carrier wavelength is $989$ nm. The fiber under investigation  has the second and third order dispersion coefficients $\beta_2 = 3.89$ ps$^2$/km and $\beta_3 = 0.0061$ ps$^3$/km at the input wavelength, and its nonlinear parameter $\gamma$ is $13$/W/km. The zero GVD wavelength is located at $\lambda = 1021$ nm, so that the pump comes in the normal GVD range and is under conditions when the interference pattern of two dispersing pulses evolves into a train of dark solitons \cite{MarestOL}. The white stripes cutting through the blue background  in Fig. \ref{f2}(c) show these solitons. The dark solitons emit their continuum of the  DWs  centered around $\simeq 1060\ $nm in the  anomalous GVD range. Spectral measurements at the propagation distance of $12\ $m, see Fig. \ref{f2}(b), show the two clear maxima corresponding to the train of dark solitons and their radiation.

A crucially important feature that we observe in our experiments and modeling is that
when the  radiation entering the range of the anomalous dispersion is sufficiently strong, as it is in the case shown in Fig. \ref{f2}, then it can spontaneously generate bright solitons \cite{bose}. A bright soliton  traverses  across the entire train of dark solitons and interacts with them and their radiation through various four-wave mixing processes. As soon as this interaction starts happening, see $z=16$m in Figs. \ref{f2}(d) and \ref{f2}(e), the continuum starts to acquire the weak but pronounced wings at its short and long wavelength edges. These wings expand much faster than the bulk of the spectrum. As soon as the bright soliton separates from the dark solitons, then the wings in question  fade away, see Figs. \ref{f2}(c) and \ref{f2}(e). When several bright solitons are generated, then we have observed the appearance of the continuum wings at every propagation distance when   these solitons were shaped with the dark soliton pattern. Note, that 
as we approach the visible spectrum the noise level is higher so that the measurements of the short wavelength continuum wing are obscured. Note, also that the short wavelength wing is visible  in the modeling for $17$m$<z<35$m, while the experimental fiber length was only $20$m.

\begin{figure}
\centering
\includegraphics[width=.49\textwidth]{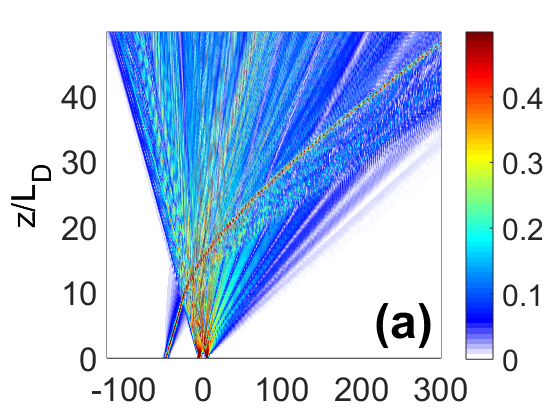}
\includegraphics[width=.49\textwidth]{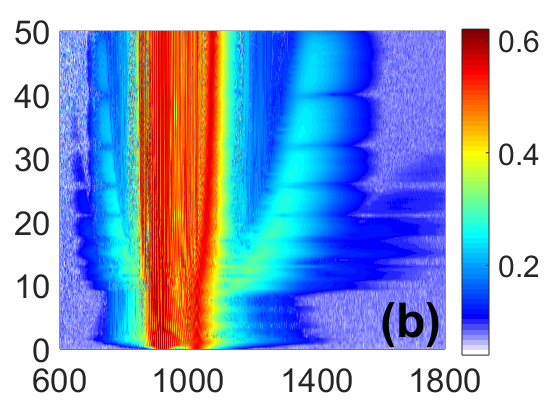}
\includegraphics[width=.49\textwidth]{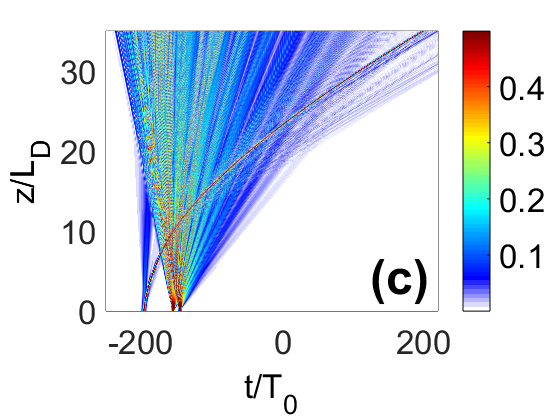}
\includegraphics[width=.49\textwidth]{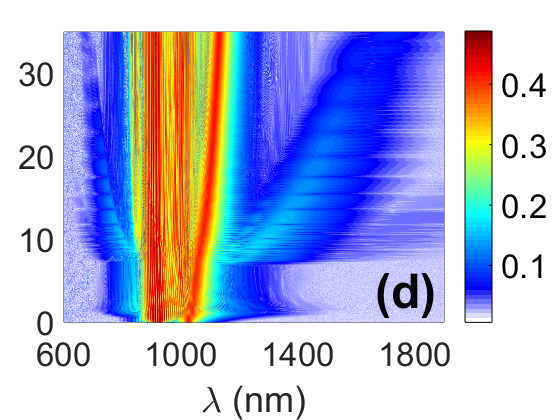}
\caption{Numerical simulation of the collision of a bright soliton with the dark soliton train. Top row disregards and the bottom one accounts for the Raman effect. (a,c) show temporal and (b,d) spectral dynamics.  Input conditions and fiber parameters as in  Fig. \ref{f2}. The spectra corresponding to the Raman shifting bright soliton and its trapped radiation, as per Ref. \cite{SkryabinNP}, correspond to the right most red  and to the left most yellow/orange stripes in (d). The diverging pale blue wings on both sides of the continuum in (b,d) is the effect studied here.
\label{f3}}
\end{figure}
To study the origin of the generation of these spectral wings, which to the best of our knowledge have not been previously noted and understood, we have carried out a series of numerical experiments under the idealized initial conditions, when the fiber is excited by an  isolated bright soliton and by the dark soliton train. The latter is arranged as in Fig. \ref{f2}, but $P_0$ is kept smaller so that the  Cherenkov continuum emitted by the dark solitons is not intense enough to spontaneously generate bright solitons. Thus,  the total  field consists of three main parts: a bright soliton, $\Psi_{bs}$, a train of dark solitons $\Psi_{ds}$ and the DWs emitted by the dark solitons $\Psi_{dsdw}$. If $g$ is the field pattern corresponding to the spectrum generated through the mixture of these three fields, then the total field $A$ takes the form:
\begin{equation}
A=\Psi+g,~\Psi=\Psi_{bs}+\Psi_{ds}+\Psi_{dsdw}.  \label{eq:ansatz}
\end{equation} 
Substituting Eq. (\ref{eq:ansatz}) into Eq. (\ref{eq:fiber}) with $Q=0$ and assuming that all three initial fields are quasi-static, we derive a propagation equation for the newly generated field $g$:
\begin{eqnarray}
i\partial_zg+\hat{D}g+2\gamma|\Psi|^2g+\gamma\Psi^2g^*=-\gamma S_{fwm}-\gamma S_{disp}\label{eq:g}.
\end{eqnarray}
The right-hand side terms in Eq. (\ref{eq:g}) are associated with the nonlinear and linear sources of the radiation field $g$. The nonlinear source $S_{fwm}$ originates in the four-wave mixing of the different combinations of the input fields \cite{SkryabinRMP}:
\begin{eqnarray}
&& S_{fwm}=\Psi_{bs}^2\Psi_{ds}^*+\Psi_{ds}^2\Psi_{bs}^*+\Psi_{dsdw}^*(\Psi_{bs}+\Psi_{ds})^2+4\Psi_{dsdw}\mathrm{Re}(\Psi_{ds}\Psi_{bs}^*)+\dots,\label{eq:SI}
\end{eqnarray}
In the above expression for $S_{fwm}$ we have disregarded all the terms  nonlinear in $\Psi_{dsdw}$ since this field is assumed much weaker than the other two. $S_{disp}$ part of the radiation source comes from the  higher order dispersion operators acting on the bright  soliton $\Psi_{bs}$ and is associated with the classic soliton Cherenkov radiation \cite{SkryabinRMP}. The terms serving as sources of the DW continuum generated by the dark solitons do not enter Eq. (\ref{eq:SI}), since this continuum is assumed to be shaped
at the earlier stages of propagation and directly enters the anzats (\ref{eq:ansatz}).
%
\begin{figure}
\centering
\includegraphics[width=.49\textwidth]{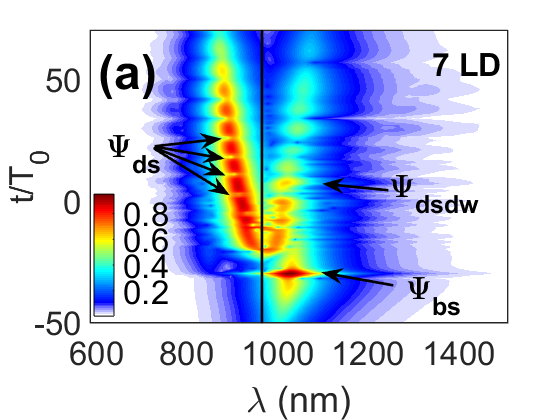}
\includegraphics[width=.49\textwidth]{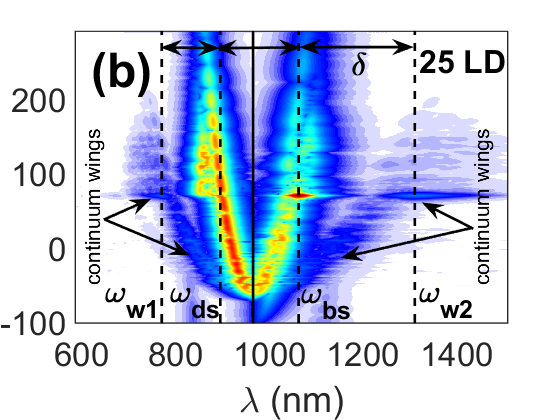}
\caption{XFROG spectrograms showing time-frequency structure  of the supercontinuum field
from Fig.\ref{f3}(a,b) for $z=7L_D$ (a) and $25L_D$ (b). Vertical solid line marks the zero  GVD wavelength. Dashed vertical lines in (b) mark the career wavelengths of the bright ($1060$nm) and dark ($895$nm) solitons, and the  wavelengths  of the continuum wings  ($770$ and $1310$nm), calculated using  Eqs. (\ref{fr}). The intense signal just on the left from the $\omega_{ds}$ line in (b) is the Cherenkov radiation emitted by the 
bright soliton.
\label{f4}}
\end{figure}
Numerical modeling results corresponding to the idealized setting described above  are shown in Fig. \ref{f3} without (top row) and with (bottom row) the Raman effect accounted for.
At $z\approx10\ L_D$ the bright soliton collides with the dark soliton train and its DWs, see Fig. \ref{f3}. As a result of this collision, the already existing continuum develops low-intensity short and long wavelength spectral wings. The wings are well separated from the bulk of the continuum and  shift further away from it  with propagation. This is exactly the dynamics we have seen in the experiment when the bright soliton was generated spontaneously. Note, that the presence or absence of the Raman scattering does not play a significant role in this effect, cf. Figs. \ref{f3}(a), \ref{f3}(b) and \ref{f3}(c), \ref{f3}(d). 

At first glance, the dynamics of the  wings reminds the effect of the DW trapping by a Raman shifting soliton \cite{SkryabinRMP,SkryabinNP,kud}. However, it quickly becomes clear that we are dealing with a different, though perhaps, a related effect here. In the radiation trapping effect \cite{SkryabinNP,kud},  the supercontinuum is shaped on the short wavelength edge by the trapped DW and on the long wavelength one by a soliton, while here the bright soliton is spectrally located between the expanding DW wings of the continuum. The trapping described in \cite{SkryabinRMP,SkryabinNP,kud} critically depends on the soliton acceleration induced by the Raman effect, which is a key factor creating the trapping potential, while in our case the wings on both edges of the continuum are formed already without the Raman effect, cf. Figs. \ref{f3} (b) and (d).  Also, the DW trapping is inherently dependent on the matching of the group velocities of the soliton and DW, while in our case the soliton and the red shifting dispersive wave are even located on the same side from the zero GVD and hence they group velocities can not be matched. Leaving the bright soliton alone in the dynamics, we have
found that its own trapped and group velocity matched Cherenkov radiation  \cite{SkryabinRMP,SkryabinNP} corresponds to the short wavelength edge of the main part of the continuum located at $845-870$nm, see Figs. \ref{f3}(d) and \ref{f4}(b).

As  the next step in our analysis, we have filtered out the Cherenkov radiation field created by the dark soliton train and have found that the generation of spectral wings persists without the $\Psi_{dsdw}$ as well. It means that the leading role in the wing formation belongs to the first two terms on the right-hand side of Eqs. (\ref{eq:SI}). The
$\Psi_{bs}^2\Psi_{ds}^*$ and $\Psi_{ds}^2\Psi_{bs}^*$ radiation source terms originate from the four-wave mixing between the bright and dark solitons. The continuous wave background of the dark soliton train is expected to play a crucial role in this process since the narrow bright soliton mostly overlaps with this background in the course of its propagation.  The frequencies that are expected to be generated by the above mixing are \begin{equation}
\omega_{w1}=2\omega_{ds}-\omega_{bs}=\omega_{ds}+\delta,~\omega_{w2}=2\omega_{bs}-\omega_{ds}=\omega_{bs}-\delta,\label{fr}\end{equation}
Here the subscript '$w$' stands for the '\textit{wing}' and $\delta=\omega_{ds}-\omega_{bs}$.

In order to  identify the spectral content of different parts of the supercontinuum field, we have plotted in Fig. \ref{f4} the XFROG diagrams for the signal at the propagation corresponding to the situation before, Fig. \ref{f4}(a), and after, Fig. \ref{f4}(b), the bright soliton starts interacting with the dark soliton train. These diagrams show the function  $X(t,\omega-\omega_0)=
\vert\int_{-\infty}^\infty\mathrm{d}\tau A(t-\tau,z_0)g(\tau)e^{i\tau(\omega-\omega_0)}\vert$,
where $g(\tau)=\mathrm{sech}(\tau)$ is the gate function \cite{DudleyRMP,SkryabinRMP}. In the 'before' diagram one can clearly see the bright soliton, which does not emit any significant Cherenkov radiation of its own, and the train of dark solitons, that have already produced a strong Cherenkov continuum. It is important to note that the broad pulse in the normal GVD range that nests the dark soliton train is chirped so that its trailing (rear) end has the wavelength (frequency)
larger (smaller), than its front end. Thus, when the bright soliton starts its journey through $\Psi_{ds}$ field it traverses across this chirped background. Hence, the frequencies $\omega_{w1}$ and $\omega_{w2}$ generated at the continuum edges do not stay constant but drift in the time-frequency space away from the zero GVD point. 
In the time domain, the pulses associated with the continuum wings emanate from the location of the bright soliton, see  Fig. \ref{f4}(d). The group velocities of the wing radiation pulses are however matched one to another across the zero GVD point, and with the greater frequency span than the spectral width of the intense part of the continuum, see also Fig. \ref{f3}.
\section*{Summary}
We  have reported the spectrally diverging short and long wavelength wings of the fiber supercontinuum generated through the interaction of the dark and bright solitons. We have found that these wings appear when the Cherenkov radiation emitted by the dark solitons into the anomalous GVD range spontaneously gives birth to a bright soliton that traverses through the dark ones. Both wings are represented by the weak dispersing pulses  emitted through the four-wave interaction between the bright and dark solitons. Our experimental and numerical data are in good agreement and a more detailed analysis of the  effects reported above represents an open challenge.
\section*{Funding}
Direction G\'en\'erale de l'Armement (DGA);
Russian Foundation for Basic Research (RFBR: 16-52-150006); ITMO University through the Government of the Russian Federation (Grant 074-U01).
T. M. and A. K. acknowledge support from 
 IRCICA (USR 3380 Univ. Lille - CNRS), from the ANR TOPWAVE (ANR-13-JS04-0004) project, from the "Fonds Europ\'{e}en de D\'{e}veloppement Economique R\'{e}gional", the Labex CEMPI (ANR-11-LABX-0007) and Equipex FLUX (ANR-11-EQPX-0017) through the "Programme Investissements d'Avenir". 

\end{document}